\documentclass[12pt]{article}
\usepackage{amsfonts}
\usepackage{amssymb}
\usepackage{color}
\usepackage{graphicx}

\usepackage{pgfplots}
\usepackage{booktabs}
\usepackage{amsmath}
\usepackage{algorithm}
\usepackage{algorithmic}
\usepackage{amsfonts}
\usepackage{latexsym}
\usepackage{subfigure}
\usepackage{latexsym}
\usepackage{graphicx,epstopdf}
\usepackage{subfigure}
\usepackage{authblk}
\usepackage{caption}
\usepackage{dsfont}

\newcommand{\R}{{\cal R}}
\newcommand{\F}{{\cal F}}
\newcommand{\Sc}{{\cal S}}
\newcommand{\Pc}{{\cal P}}

\newcommand{\mc}{\mathcal}

\newcommand{\be}{\begin{equation}}
\newcommand{\en}{\end{equation}}
\newcommand{\bea}{\begin{eqnarray}}
\newcommand{\ena}{\end{eqnarray}}
\newcommand{\beano}{\begin{eqnarray*}}
\newcommand{\enano}{\end{eqnarray*}}

\newcommand{\1}{1 \!\! 1}
\newcommand{\ST}{\mc S}

\newcommand{\Hil}{\mc H}

\catcode `\@=11 \@addtoreset{equation}{section}

\catcode `\@=12
\begin{document}

\title{Modeling epidemics through ladder operators}

\author{F. Bagarello$^{1,2}$, F. Gargano$^{1}$, F. Roccati$^3$\\
	
	\small{
		$^1$Dipartimento di Ingegneria -  Universit\`a di Palermo,
		Viale delle Scienze, I--90128  Palermo, Italy,}\\
	\small{ $^2$I.N.F.N -  Sezione di Napoli, Italy}\\
	\small{
		$^3$Dipartimento di Fisica e Chimica Emilio Segr\`e, Universit\`a degli Studi di Palermo, 
		via Archirafi 36, I-90123 Palermo, Italy.}\\
	\small{\emph{Email addresses:}\\
		fabio.bagarello@unipa.it,
		francesco.gargano@unipa.it,
		federico.roccati@unipa.it}
}


\date{}
\maketitle

\begin{abstract}
\noindent We propose a simple model of spreading of some infection in an originally healthy population which is different from other models existing in the literature. In particular, we use an {\em operator technique} which allows us to describe in a natural way the possible interactions between healthy and un-healthy populations, and their transformation into recovered and to dead people.
{After a rather general discussion, we apply our method to the analysis of Chinese data for the SARS-2003 (Severe acute respiratory syndrome; SARS-CoV-1) and the Coronavirus COVID-19 (Corona Virus Disease; SARS-CoV-2 )}  and we show that the model works very well in reproducing the long-time behaviour of the disease, and in particular in finding the number of affected and dead people in the limit of large time. Moreover, we show how the model can be easily modified to consider some lockdown measure, and we deduce that this procedure drastically reduces the asymptotic value of infected individuals, as expected, and observed in real life.
\end{abstract}

\vfill

\newpage


\section{Introduction}

In recent months the entire world has faced an emergency due to the outbreak of the severe acute respiratory syndrome coronavirus 2 (SARS-CoV-2). As for other infectious diseases outbreak, it is crucial to estimate the transmission dynamics of the infection in the various stages of the disease. This can provide insights into the epidemiological situation and identify whether outbreak control measures have  significant effects, \cite{McBry2020}. Moreover, the potentiality of predicting a dynamical equilibrium in which the total amount of infected individuals has reached the plateau  is of valuable importance, as it helps to estimate the risk for other countries or  forecasts the possible conclusion of the emergency.

In this respect a mathematical (and statistical) analysis and its connection with biological aspects play a crucial role.  
This analysis can include the  statistical evaluation of the up to date relevant quantities such as the  various growth rates of the numbers of the infected, dead and recovered individuals through statistical arguments and best fit procedures with  well known phenomenological  models like the exponential and logistic one.
More sophisticated stochastic models can of course be adopted, providing more precise results: they can take into account for  time-varying extrinsic factors typical of the epidemic course (social cycles  and  climatic variations, see discussions in \cite{Caz,Sha}) and viral phylodynamics with genetic variations, see \cite{Dureau, Rasm}.
Other models can increase the reliability of the forecasting including agent-based simulations capable to take into account for a variety of complex connections and the population mobility, \cite{Eub}.
{Needless to say, a huge research activity has been devoted in the last months to the mathematical modeling related to the spreading of COVID-19, using different kind of approaches: Monte Carlo simulations, \cite{cov2}, deep learning and fuzzy rule \cite{cov3}, compartments/SIR-like epidemic models, \cite{cov1,cov4}, transmission network models, \cite{cov5}, to cite a few.}
Despite the fact that the various models can include a large variety of variables that can improve the reliability of the forecasting outcomes, the unpredictable intrinsic variability of any infection leads very often to the failure of the adopted model in predicting the evolution of the infection process. Most models, especially the very basic ones, could work only to reproduce the evolution of the infection up to the moment the model is applied, and the best parameters used to fit real data can dramatically change when the  data are upgraded. This is evident in these days, when many groups of scientists were (and are) forced to change everyday their own estimates on the final numbers of infected or dead people, or on the plausible end of the emergency.
The world of the  statistical and mathematical methods applied to  the detection of disease outbreaks is huge and constantly growing, and we refer to \cite{Unk,Sie} for  recent  reviews containing further literature and valuable comments on this field.

The common idea behind the various models is that the host
population is divided into distinct classes, according to its epidemiological status: the classical $SIR$ formalism is often used, according to the classes Susceptible to the
disease ($S$), currently Infectious ($I$), and Recovered ($R$), and adding other classes (for instance Exposed ($E$) or Dead ($D$)) is commonly done to give depth to the model.
In this paper we adopt an operatorial method based on certain ladder operators appearing often in quantum mechanical systems (but not only)  to describe the long time dynamics of an infected population, assuming that the main classes of the model are the  infected, recovered and dead individuals. As we shall see, the susceptible/healthy individuals play  the role of ``reservoir'' for the infected people. The underlying idea of our model is to construct
an Hamiltonian operator $H$ for a certain system $\mathcal S$ using ladder operators, and using the Heisenberg equations of motion to deduce the time evolution of some relevant observables of $\mathcal S$, the aforementioned classes of individuals, on a macro-scale (we actually compute the time evolution of the densities of these classes). 
The Hamiltonian $H$  contains all the interactions occurring between
the different agents of $\mathcal S$. In particular, ladder operators can be easily used to describe the mechanisms leading to the infection process and to the subsequent recovery or death, and to define a suitable $H$ giving rise to these processes. 

This is actually the first attempt to use operatorial methods  in application to infection processes: they have been successfully used in various biological contexts, to fit real data  by modeling population-resources dynamics, \cite{Garg}, to model basic cancer cell dynamics, \cite{BagEntr,Rob},  to reproduce biological aspects of the bacterial dynamics, \cite{DOLI}, and to model a basic epigenetic
evolution, \cite{Asa} (see \cite{bagbook1,bagbook2} for other fields of application). Our main goal is  to provide a reliable long time dynamics  capable to capture the final stage of the infection when the various densities of the classes have reached some sort of equilibrium values. For concreteness, we shall apply our model to the the recent COVID 19 outbreaks and to the previous SARS in 2003.

\vspace{2mm}

The paper is organized as follows: in Section \ref{sec::sec2} we present the model and the mathematical settings, and we define the main operators used to construct the Hamiltonian of the system. We derive the dynamics in Section \ref{sec::sec22}. In Section \ref{sec::sec3} we apply our model to capture the long time behavior of the recent Coronavirus pandemics in China and in the Italian region of Umbria, and that of the SARS epidemics in 2003 in China. Furthermore, by slightly enriching our model, we show how imposing lockdown measures after 10 days from the beginning of the spreading of the epidemics is beneficial for the long time value of infected, recovered and dead individuals.
Our conclusions are given in Section \ref{sec:conclo}, while the Appendices contain few facts on the number representation of quantum mechanics (Appendix 1) and some useful formula on the dynamics of the populations (Appendix 2).

\section{The model and its dynamics}\label{sec::sec2}

In this section we will introduce our model, starting with a general description of its components, introducing then to the possible interactions between them leading to the  Hamiltonian of the system, and deducing, out of it, the differential equations of motion and their solutions, with particular interest to their long time behavior. We refer to \cite{bagbook1} and to \cite{bagbook2} for many details on this approach,  for many different applications of this strategy, and for many references.

\begin{figure}[!h]\centering

	\tikzset{every picture/.style={line width=0.75pt}} 
	
	\begin{tikzpicture}[x=0.75pt,y=0.75pt,yscale=-1,xscale=1]
	
	\draw   (106,76.84) .. controls (106,72.51) and (109.51,69) .. (113.84,69) -- (182.35,69) .. controls (186.68,69) and (190.19,72.51) .. (190.19,76.84) -- (190.19,100.34) .. controls (190.19,104.67) and (186.68,108.18) .. (182.35,108.18) -- (113.84,108.18) .. controls (109.51,108.18) and (106,104.67) .. (106,100.34) -- cycle ;
	\draw   (234,58) .. controls (234,53.58) and (237.58,50) .. (242,50) -- (296,50) .. controls (300.42,50) and (304,53.58) .. (304,58) -- (304,82) .. controls (304,86.42) and (300.42,90) .. (296,90) -- (242,90) .. controls (237.58,90) and (234,86.42) .. (234,82) -- cycle ;
	\draw   (348,79) .. controls (348,74.58) and (351.58,71) .. (356,71) -- (410,71) .. controls (414.42,71) and (418,74.58) .. (418,79) -- (418,103) .. controls (418,107.42) and (414.42,111) .. (410,111) -- (356,111) .. controls (351.58,111) and (348,107.42) .. (348,103) -- cycle ;
	\draw   (209.19,209.39) .. controls (209.19,202.55) and (214.73,197) .. (221.58,197) -- (333.8,197) .. controls (340.64,197) and (346.19,202.55) .. (346.19,209.39) -- (346.19,246.55) .. controls (346.19,253.39) and (340.64,258.94) .. (333.8,258.94) -- (221.58,258.94) .. controls (214.73,258.94) and (209.19,253.39) .. (209.19,246.55) -- cycle ;
	\draw    (189.19,114.18) -- (234.64,189.61) ;
	\draw [shift={(236.19,192.18)}, rotate = 238.93] [fill={rgb, 255:red, 0; green, 0; blue, 0 }  ][line width=0.08]  [draw opacity=0] (8.93,-4.29) -- (0,0) -- (8.93,4.29) -- cycle    ;
	\draw    (270.16,191) -- (269.22,102.18) ;
	\draw [shift={(269.19,99.18)}, rotate = 449.39] [fill={rgb, 255:red, 0; green, 0; blue, 0 }  ][line width=0.08]  [draw opacity=0] (8.93,-4.29) -- (0,0) -- (8.93,4.29) -- cycle    ;
	\draw    (197.97,85.05) -- (229.41,72.31) ;
	\draw [shift={(232.19,71.18)}, rotate = 517.9300000000001] [fill={rgb, 255:red, 0; green, 0; blue, 0 }  ][line width=0.08]  [draw opacity=0] (8.93,-4.29) -- (0,0) -- (8.93,4.29) -- cycle    ;
	\draw [shift={(195.19,86.18)}, rotate = 337.93] [fill={rgb, 255:red, 0; green, 0; blue, 0 }  ][line width=0.08]  [draw opacity=0] (8.93,-4.29) -- (0,0) -- (8.93,4.29) -- cycle    ;
	\draw    (311.87,73.52) -- (340.5,87.84) ;
	\draw [shift={(343.19,89.18)}, rotate = 206.57] [fill={rgb, 255:red, 0; green, 0; blue, 0 }  ][line width=0.08]  [draw opacity=0] (8.93,-4.29) -- (0,0) -- (8.93,4.29) -- cycle    ;
	\draw [shift={(309.19,72.18)}, rotate = 26.57] [fill={rgb, 255:red, 0; green, 0; blue, 0 }  ][line width=0.08]  [draw opacity=0] (8.93,-4.29) -- (0,0) -- (8.93,4.29) -- cycle    ;
	\draw   (68,86.94) .. controls (68,54.35) and (160.27,27.94) .. (274.09,27.94) .. controls (387.92,27.94) and (480.19,54.35) .. (480.19,86.94) .. controls (480.19,119.52) and (387.92,145.94) .. (274.09,145.94) .. controls (160.27,145.94) and (68,119.52) .. (68,86.94) -- cycle ;
	
	\draw (108,79.84) node [anchor=north west][inner sep=0.75pt]   [align=left] {Recovered};
	\draw (241,63) node [anchor=north west][inner sep=0.75pt]   [align=left] {Infected};
	\draw (363,82) node [anchor=north west][inner sep=0.75pt]   [align=left] {Dead};
	\draw (233,211) node [anchor=north west][inner sep=0.75pt]   [align=left] {Healthy};
	\draw (434,71) node [anchor=north west][inner sep=0.75pt]  [font=\large] [align=left] {$\displaystyle \mathcal{S}_{\mathcal{P}}$};
	\draw (315,226) node [anchor=north west][inner sep=0.75pt]  [font=\large] [align=left] {$\displaystyle \mathcal{R}$};

	\end{tikzpicture}\caption{\label{figscheme} The system and its multi-component reservoir.}
	
\end{figure}
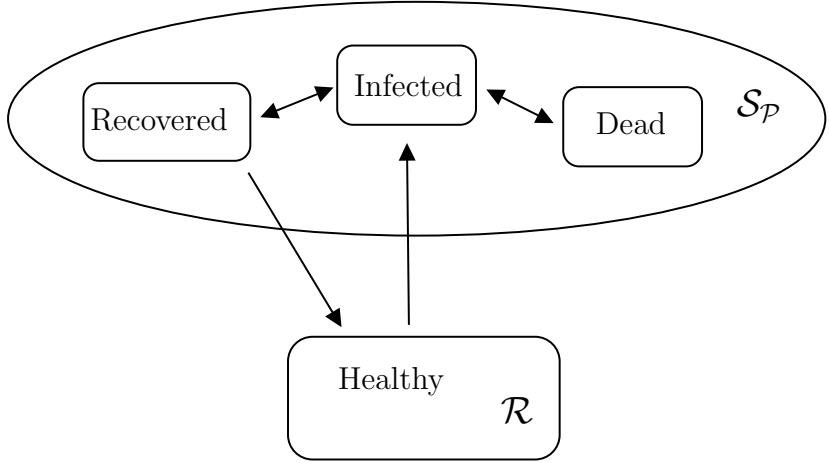

Our system $\Sc$ consists of four different compartments:  {\em healthy},  {\em infected},  {\em recovered} and  {\em dead} individuals. These populations, when considered as sets, {\em are not of the same  kind}: in particular, the set of healthy people is very large, when compared to the other three. This is because a small percentage of the whole population is infected, in most epidemics experienced so far in mankind's history. For this reason, using a well settled approach, we consider the healthy people as a sort of big {\em reservoir}, $\R$, for a smaller (in size) system, $\Sc_\Pc$, made of the other three compartments. Then $\ST=\R\cup\Sc_\Pc$. These compartments interact as diagrammatically shown in Figure \ref{figscheme}: first of all, healthy people can be infected. Those who have been infected, either die or   recover. The latter fill up again the set of healthy people. These are the essential reasonable mechanisms which we imagine occur during an epidemic, according to the literature on this topic. We are not going into the biological reasons why the contagion can spread all along the population. We only want to describe  the possible consequences of our simple view to the system, and show how the long time behaviour of the number of the three relevant populations here can be recovered, and are in good agreement in some concrete applications of our general model.

 \vspace*{1cm}

As in similar approaches on several dynamical systems considered using operator techniques, see \cite{bagbook1,bagbook2}, we attach some ladder operators to each compartment of the system, and we assume these operators obey suitable commutation relations. The particular choice of these relations is very much related to the characteristics of the system we are interested in. Here we will assume that the relevant rules are the {\em anti-commutation relations}, (CAR), as we propose next. The reason for this choice will be clarified later.

The ladder operators for the infected are $p_1$ and $p_1^\dagger$, those for the recovered are $p_2$ and $p_2^\dagger$, while $p_3$ and $p_3^\dagger$ are those for the dead. To the healthy people is attached  a family of ladder operators, labeled by a continuous index $k\in\mathbb{R}$, $B(k)$ and $B^\dagger(k)$.  The following rules are assumed:
\be\{p_n,p_m^\dagger\}=\delta_{n,m}\1, \quad \{B(k),B^\dagger(q)\}=\delta(k-q)\,\1,
\label{21}\en
for all $n,m=1,2,3$ and  $k,q\in\mathbb{R}$. Here $\{x,y\}:=xy+yx$. All the other anti-commutators are taken to be zero. For instance,
\be
\{p_n,p_m\}=\{B(k),B(q)\}=\{p_n,B(k)\}=\{p_n,B^\dagger(k)\}=\ldots=0.
\label{22}
\en
The rationale for this choice, widely discussed in recent years, see \cite{bagbook1,bagbook2} and references therein, is mainly based on the simple remark that the use of ladder operators is an easy and natural way to describe exchange mechanisms between different compartments of a composite system. This aspect will be clarified soon. 

The operators above are used to construct an operator, the Hamiltonian $H$ of $\Sc$, which is the generator of the dynamics of the system. In details, we will assume the following expression for $H$:
\be
\left\{
\begin{array}{ll}
	H=H_{0}+H_{1}+H_{2}, &  \\
	H_{0}=\sum_{j=1}^{3}\omega _{j}p_j^\dagger p_j+\int_{\Bbb R}\Omega(k)B^\dagger(k)B(k)\,dk,   \\
	H_{1}=\sigma_{B,1}\int_{\Bbb R}\left(p_1 B^\dagger(k)+B(k)p_1^\dagger\right)\,dk+\sigma_{B,2}\int_{\Bbb R}\left(p_2 B^\dagger(k)+B(k)p_2^\dagger\right)\,dk,\\
	H_{2}=\sigma_{1,2}\left(p_1^\dagger p_2+p_2^\dagger p_1\right)+\sigma_{1,3}\left(p_1^\dagger p_3+p_3^\dagger p_1\right).
\end{array}%
\right.
\label{23}\en
Here $\omega _{j}$ and the $\sigma$'s are all real quantities, and $\Omega(k)$ is a real-valued function, so that $H$ appears as a (formally) Hermitian operator.
The meaning of the various contributions in $H$, and their relations with the scheme in Figure \ref{figscheme}, is quite standard in the operatorial interpretation of an Hamiltonian.
The first term in $H_1$ models the vertical line in Figure \ref{figscheme}. In particular, $B(k)p_1^\dagger$ describes the fact the number of healthy people decrease (because of $B(k)$, which is a lowering operator), while the density of the infected increases because of $p_1^\dagger$, which is a raising operator\footnote{Of course, a full comprehension of this mechanism is possibly having a knowledge of the functional Hilbert space structure connected to CAR. In particular, the action of creation and annihilation operators on certain vectors is the crucial ingredient in this construction. All these details, which are well known to all who are familiar with second quantization, can be found in \cite{roman}, or in \cite{bagbook1,bagbook2} more in connection with what is relevant for this paper.  We also list some useful results in Appendix 1}. Notice that $H_1$ also contains the adjoint term $p_1 B^\dagger(k)$, which can be seen as a recovery term: the density of infected decreases (as the effect of $p_1$), while that of healthy people increases, because of $B^\dagger(k)$. Our model also admits a two-steps recovery path: first the infected slightly recover, and then they recover completely. It is like going from the intensive care to other departments in the hospital, and then being released. This double steps are described by $\sigma_{1,2}\left(p_1^\dagger p_2+p_2^\dagger p_1\right)$ in $H_2$, and by the second term in $H_1$. As before, we have a contribution $p_2 B^\dagger(k)$ describing full recovery, but we also have a term $B(k)p_2^\dagger$ which describes the possibility that someone, negative to a first swab, becomes positive later. This term could also be relevant in describing people who fall again infected, after a first recovery. A similar interpretation we can imagine for the term  $\sigma_{1,2}p_1^\dagger p_2$: part of those which apparently are recovering, fall again infected. As for $H_0$, this is a standard {\em free} term, not affecting all the populations in absence of interactions. In other words, if $H=H_0$, the densities of the members of each population does not change with time. However, in all the systems considered along the years, $H_0$ becomes relevant mainly in presence of interactions, and describes some inertia of the various compartments, see \cite{bagbook1,bagbook2}.

It is clear that the analytic form of $H$ could be enriched, for instance by adding extra terms, by requiring some explicit time dependence in the Hamiltonian itself, or also by relaxing the assumption that $H$ should be Hermitian, \cite{fg,fb2020}. We could also refine our analysis by assuming that some external rule acts on the system, as proposed in \cite{fffr}. This possibility will be considered later on, and we will show that it produces interesting results.

\vspace{1mm}

We can now go back to the analysis of the dynamics of the system, by assuming that this is driven by the same standard mechanisms which work well in presence of operator-valued unknown. In particular, we use the Heisenberg equations of motion $\dot X(t)=i[H,X(t)]$, \cite{bagbook1,bagbook2}, which produces, by using the CAR (\ref{21}) and (\ref{22}) above, the following:

\be
\left\{
\begin{array}{ll}
	\dot p_1(t)=-i\omega_1 p_1(t)+i\sigma_{B,1}\int_{\Bbb R}B(q,t)\,dq-i\sigma_{1,2}p_2(t)-i\sigma_{1,3}p_3(t),\\
	\vspace{1mm}
	\dot p_2(t)=-i\omega_2 p_2(t)+i\sigma_{B,2}\int_{\Bbb R}B(q,t)\,dq-i\sigma_{1,2}p_1(t),\\
	\vspace{1mm}
	\dot p_3(t)=-i\omega_3 p_3(t)+i\sigma_{1,3}p_1(t),\\
	\vspace{1mm}
	\dot B(q,t)=-i\Omega(q) B(q,t)+i \sigma_{B,1} p_1(t)+i \sigma_{B,2} p_2(t).   \label{24}
\end{array}%
\right.
\en
This is a system of four ordinary, operator valued, linear differential equations in four unknowns which describes the time evolution of the essential dynamical variables of the system, the lowering operators $p_j(t)$, $j=1,2,3$ and $B(q,t)$. So it is not surprising that we can deduce an explicit analytic (but rather implicit) solution. As a matter of fact, this system is a simplified version of the one deduced in \cite{bagpol}, to which we refer for the details of the solution of this system. Here we briefly review some essential steps, focusing on what is interesting for us.

{From the time evolution of the above operators, we can deduce the time evolution of the relevant  operators  used to measure the densities of the various populations. In fact we consider  the evolution of the {\em number operators} related to the $p_j(t)$ above, $\hat P_j(t)=p_j^\dagger(t)p_j(t)$, $j=1,2,3$, and then to their mean values on suitable states describing the initial conditions of $\Sc$, see below, i.e., for us, the initial numbers (or, more properly, densities) of the infected, recovered and dead. It is very well known that the mean values of the number operators attached to ladder operators satisfying the CAR are always quantities ranging in $[0,1]$, in agreement with our interpretation of these quantities as  densities of the populations of $\Sc_\Pc$ from a macroscopic point of view.}

\subsection{The state on $\Sc$}\label{sec::sec21}

Because of the relevance of the CAR for our model, it is natural to introduce the vacuum of the $p_j$, i.e. the non zero vector $\varphi_{0,0,0}$ in $\Hil_\Pc=\mathbb{C}^8$, the Hilbert space of $\Sc_\Pc$, satisfying the condition $p_j\varphi_{0,0,0}=0$, $j=1,2,3$. The other vectors of the orthonormal basis $\F_\varphi=\{\varphi_{n_1,n_2,n_3},\,n_1,n_2,n_3=0,1\}$ for $\Hil_\Pc$ can be constructed as follows:
$$
\varphi_{1,0,0}=p_1^\dagger\varphi_{0,0,0}, \quad \varphi_{0,1,0}=p_2^\dagger\varphi_{0,0,0}, \quad \varphi_{1,1,0}=p_1^\dagger\,p_2^\dagger\varphi_{0,0,0},\quad \varphi_{1,1,1}=p_1^\dagger\,p_2^\dagger\,p_3^\dagger\varphi_{0,0,0},
$$
and so on. Now, the reason why $\hat P_j(t)$ is called "number operator" is because, if we compute is action on $\varphi_{n_1,n_2,n_3}$, we get $\hat P_j\varphi_{n_1,n_2,n_3}=n_j\varphi_{n_1,n_2,n_3}$, $j=1,2,3$ and $\hat P_j=\hat P_j(0)$. This means that the elements of $\F_\varphi$ are eigenstates of $\hat P_j$, properties which is lost for $t>0$: in general,  $\hat P_j(t)\varphi_{n_1,n_2,n_3}$ is not even proportional to $\varphi_{n_1,n_2,n_3}$.

In view of what we have discussed before, the subsystem $\Sc_\Pc$ is described, at $t=0$, by the vector $\varphi_{0,0,0}$. This is because the eigenvalues of the $\hat P_j(t)$'s, or better (see below), their mean values, are proportional to the densities of the various compartments. Therefore, since at $t=0$ there are no elements of $\Pc_1$, $\Pc_2$ and $\Pc_3$, $\Sc_\Pc$ must be described by $\varphi_{0,0,0}$, which is the only vector in $\F_\varphi$ with eigenvalues $n_1=n_2=n_3=0$.

At this stage, we observe that $\Sc_\Pc$ is just a part of $\Sc$: we still have to clarify how a state over $\R$ should be defined, and which are its properties. Once again, we refer to \cite{bagbook1,bagbook2} for the details: here we only stress that the state over $\R$ is not expected to be of the same kind of that for $\Sc_\Pc$, because these two parts of $\Sc$ are really different in size: $\R$ is enormous, while each $\Pc_j$ is relatively small. This is reflected by the fact that, while $\Sc_\Pc$ is described in terms of just 3 operators acting on an 8-dimensional Hilbert space, the full description of $\R$ requires a continuous number of operators\footnote{We could use also a discrete, but still infinite, number of ladder operators, but this would make many computations more complicated, \cite{bagbook1,bagbook2}.}. The state on $\R$ is a positive linear functional, \cite{rs,br}, on the algebra of the reservoir operators satisfying the following (standard) rules:
 \be
 \omega _{\R}(1\!\!1_{\R})=1,\quad \omega _{\R}(B(k))=\omega
 _{\R}(B^{\dagger }(k))=0,\quad \omega _{\R}(B^{\dagger
 }(k)B(q))=N(k)\,\delta (k-q),
 \label{25}\en
 for some suitable function $N(k)$, as well as $\omega
 _{\R}(B(k)B(q))=0$.
{In analogy with what happens for $\Sc_\Pc$, $N(k)$ is also set to be constant in $k$. From now on, we set $N(k)=N$, and this value is a kind of measure of the density of people exposed to the infection, i.e. a way to measure the strength of the infection process: $N=0$ actually means that there are no people exposed  (or in other way absence of infection), whereas increasing value of $N$ gives strength to the infection (i.e. more people are exposed). This will be clear later on when we  show a specific asymptotic solution corresponding to a particular choice of the parameters, and changing the value of $N$ can drastically modify the whole dynamics, an aspect we shall use in our applications.}

 The state over the full system $\Sc$, $\left<.\right>$, is defined, for all operators of the form $X_{\Sc}\otimes Y_{\R}$, $X_{\Sc}$ being  an operator of $\Sc_\Pc$ and $Y_{\R}$ an operator of
 the reservoir, as follows: 
\be
\left\langle X_{\Sc}\otimes Y_{\R}\right\rangle :=\left\langle \varphi_{n_1,n_2,n_3},X_{\Sc}\,\varphi_{n_1,n_2,n_3}\right\rangle \,\omega
_{\R}(Y_{\R}).
\label{26}\en
For  the purposes of our model, we define  the mean values of the number operators, which are what we can call {\em the density functions} for the system  \be P_j(t):=\left<\hat P_j(t)\right>=\left<p_j^\dagger(t)p_j(t)\right>,\label{27}\en $j=1,2,3$. Before going back to the analysis of the dynamics, it might be useful to observe that definition (\ref{27}) implies that, taking $\varphi_{n_1,n_2,n_3}=\varphi_{0,0,0}$ in (\ref{26}), $P_j(0)=0$ for $j=1,2,3$. This is because $\hat P_j(0)$ is identified with $\hat P_j(0)\otimes \1_\R$, recalling that $\omega _{\R}(1\!\!1_{\R})=1$. This result obviously reflects the initial conditions assumed for $\Sc_\Pc$.

\subsection{Back to the system}\label{sec::sec22}

The details of the computations of $P_j(t)$ can be found, for a different (and slightly larger) system, in \cite{bagbook2,bagpol}. Here we only write the final analytical expression for $P_j(t)$ in (\ref{27}). For that, we need to introduce first some useful quantities. We start defining the symmetric block matrix $U$ as follows
$$U=\left(
\begin{array}{cccccc}
\hat\omega_1 & \gamma_{1,2} & \gamma_{1,3}  & 0 & 0 & 0 \\
\gamma_{1,2} & \hat\omega_2 & 0 & 0 & 0 & 0 \\
\gamma_{1,3} & 0 & \hat\omega_3 & 0 & 0 & 0 \\
0 & 0 & 0 & \overline{\hat\omega_1} & \overline{\gamma_{1,2}} & \overline{\gamma_{1,3}} \\
0 & 0 & 0 & \overline{\gamma_{1,2}} & \overline{\hat\omega_2} & 0 \\
0 & 0 & 0 & \overline{\gamma_{1,3}} & 0 & \overline{\hat\omega_3} \\
\end{array}
\right),
$$
where
$$
\hat\omega_1:=i\omega_1+\pi\frac{\sigma_{B,1}^2}{\Omega}, \quad \hat\omega_2:=i\omega_2+\pi\frac{\sigma_{B,2}^2}{\Omega}, \quad \hat\omega_3:=i\omega_3,  \quad  \gamma_{1,2}:=i\sigma_{1,2}+\frac{\pi}{\Omega}\sigma_{B,1}\sigma_{B,2}, \quad  \gamma_{1,3}:=i\sigma_{1,3}.
$$
Here $\Omega$ is a positive constant related to $\Omega(k)$, which we assume is linear in $k$:  $\Omega(k)=\Omega k$. Further we call $V_t:=e^{-Ut}$,  $(V_t)_{j,k}$ its $(j,k)$-th matrix element, and 
$$
p_k^{(j)}(t)=\left|(V_t)_{j,k}\right|^2, \qquad p_{k,l}^{(j)}(t)=2\Re\left[\overline{(V_t)_{j,k}}\,(V_t)_{j,l}\right],
$$
where $\Re(z)$ stands for the real part of the complex quantity $z$. 
We write the explicit formula for the density functions assuming first that the initial state on $\Sc_{\cal P}$ coincides with one of the element of $\mathcal{F}_\varphi$ and then extending the result to the general case in which this state is a linear combination (normalized to one) of the elements of $\mathcal{F}_\varphi$.
If the initial state is $\varphi_{n_1,n_2,n_3}$ we have
\be
P_j(t)=P_j^{(a)}(t)+P_j^{(b)}(t),
\label{28}\en
where
\be
P_j^{(a)}(t)=\sum_{k=1}^3\left(\left|(V_t)_{j,k}\right|^2n_k+\left|(V_t)_{j,k+3}\right|^2(1-n_k)\right)
\label{29}\en
and
\be
P_j^{(b)}(t)= \frac{2\pi N}{\Omega}\int_0^tdt_1\Pi_j(t-t_1),
\label{210}\en
$j=1,2,3$, where, following \cite{bagbook2,bagpol},
\be
\Pi_j(s)=\left|\sigma_{B,1}(V_s)_{j,1}+\sigma_{B,2}(V_s)_{j,2}\right|^2.
\label{211}\en

In particular, let us see what happens if we assume that, at $t=0$, the three $\Pc_j$ populations are empty, while all the people is healthy (which is what we have already assumed several times so far). In other words: the infection starts to propagate at $t=0$ from a situation in which everyone is healthy. This fixes uniquely the state on which the mean values of the  $\hat P_j(t)$ must be computed. If this is the case, then the values of the $n_k$ in (\ref{29}) are all zero. Moreover, the block form of the matrix $U$ is translated to a similar form for $V_t$. This implies that $(V_t)_{j,k+3}=0$ for all $j,k=1,2,3$. Hence the conclusion is that,
assuming that at $t=0$ the number of infected, recovered and dead is zero, the term $P_j^{(a)}(t)$ does not contribute to $P_j(t)$ in (\ref{28}).
%
%
It is possible in this case to determine the asymptotic limit of the density functions :
\be
P_j(\infty)=\lim_{t,\infty} \frac{2\pi N}{\Omega}\int_0^\infty dt_1\Pi_j(t_1),
\label{212}\en
which are directly related to the long time behaviour of the various compartments in a {\em real} epidemic. { As anticipated, the value of $N$ modifies the asymptotic value of the density functions, and increasing  $N$ increases the infected and as a consequence the recovered and the dead, whereas if $N=0$ no one is infected.}
For a specific choice of parameters, the limits in eq.~\eqref{212} can be analytically computed. Indeed, setting $\sigma_{B,2}=0$ and $\omega_{j}=0$ $(j=1,2,3)$ we get, as deduced from the results in the Appendix 2,
\begin{equation}\label{limits}
P_1(\infty) = N,
\quad 
P_2(\infty) = \frac{N \sigma_{1,2}^2}{\sigma_{1,3}^2 + \sigma_{1,2}^2}, 
\quad 
P_3(\infty) = \frac{N \sigma_{1,3}^2}{\sigma_{1,3}^2 + \sigma_{1,2}^2}. 
\end{equation} 

Let us now see what happens if the state on $\Sc_\Pc$ is of the form 
\bea\label{214b}
\Psi=\sum_{n_1,n_2,n_3=0,1}\alpha_{n_1,n_2,n_3}\varphi_{n_1,n_2,n_3},\, \textrm{ with } \sum_{n_1,n_2,n_3=0,1}|\alpha_{n_1,n_2,n_3}|^2=1.
\ena
This extension is relevant for what discussed in Section \ref{sectlockmea}, since it will not be possible, in that case, assume that the system starts its evolution from a single $\varphi_{n_1,n_2,n_3}$. In this case the densities are given for all $j$ by
 
 \begin{equation}
 P_{j}(t)=P^{(a)}_{j}(t)+\delta P^{(a)}_{j}(t)+P^{(b)}_{j}(t),\qquad j=1,2,3,
 \end{equation}
where

\begin{equation*}\begin{aligned}
P^{(a)}_{j}(t)=\left(\sum_{n_2,n_3=0,1}\left|(V_t)_{j,1}\right|^{2}\left|\alpha_{1,n_2,n_3}\right|^{2}\right)+\left(\sum_{n_1,n_3=0,1}\left|(V_t)_{j,2}\right|^{2}\left|\alpha_{n_1,1,n_3}\right|^{2}\right)+\\
\left(\sum_{n_1,n_2=0,1}\left|(V_t)_{j,3}\right|^{2}\left|\alpha_{n_1,n_2,1}\right|^{2}\right), \\
\delta P^{(a)}_{j}(t)= 2 \Re\left[\overline{(V_t)_{j,1}} (V_t)_{j,2} \left(\sum_{n_3=0,1}\bar{\alpha}_{1,0,n_3} \alpha_{0,1,n_3} \right)+
\overline{(V_t)_{j,1}} (V_t)_{j,3} \left(\sum_{n_2=0,1}\bar{\alpha}_{1,n_2,0} \alpha_{0,n_2,1} \right)+\right.\\
\left. \overline{(V_t)_{j,2}} (V_t)_{j,3} \left(\sum_{n_1=0,1}\bar{\alpha}_{n_1,1,0} \alpha_{n_1,0,1} \right) \right]
\end{aligned}\end{equation*}
and $
P^{(b)}_j(t)$ is the same as in \eqref{210}.
We observe  that $P^{(a)}_j$ reduces to \eqref{29} if the initial state coincides with (only) one of the $\varphi_{n_1,n_2,n_3}$ and $\delta P^{(a)}_j(t)=0$. Otherwise $\delta P^{(a)}_j(t)\neq0$, and it can be considered as an \textit{interference} term which appears only if the state  is a non trivial superposition of the elements of $\mathcal{F}_{\varphi}$. We will show that this may produce some not negligible oscillations of the $P_j(t)$, which might appear strange. However, this is nor really a big surprise: similar oscillations have been observed, in a different context but with a similar approach, when dealing with decision making, \cite{bhk}. This happens again in presence of non trivial superpositions of states, as in formula (\ref{214b}). 
 These oscillations are particularly evident during the first part of the time evolution, while they tend to disappear for $t$ sufficiently large. The mathematical source of the oscillations is  the Hermiticity of $H$, and could be possibly avoided by removing this constraint, as it was done in \cite{BagEntr} in a different biological application of ladder operators. Here we are not particularly worried by these oscillations, since our main interest is in the long-time dynamics of the system, where in any case the oscillations disappear (or are negligible).


\section{Applications}\label{sec::sec3}

In this section we show how this model can efficiently reproduce the long time dynamics in some recent epidemics. In particular, we will concentrate on SARS and Coronavirus in China, and on Coronavirus in the Italian region of Umbria.

Coronavirus pandemic, as far as we know today, started in Wuhan, China, most likely at the end of December 2019. Until January 2020 the epidemic was apparently confined to China with few thousands of confirmed cases, while between February and March it spread in Europe, in the US, and in other countries,  being therefore declared a pandemic and reaching a total number of confirmed cases of more than eight million, still growing.

The 2002-2004 severe acute respiratory syndrome SARS outbreak was first identified in Foshan, Guangdong, China, in late 2002. Up to May 2004 it caused over 8,000 individual infections from 29 different countries, and more than 900 deaths worldwide. The major part of the outbreak was concentrated in the first part of the 2003 despite several SARS cases were reported until May 2004.

We show here the results concerning the long time behavior by capturing trough best-fit procedures the parameters which minimize the relative error with the precise values of cumulative infected, recovered and dead individuals in China for both SARS and Coronavirus and in Umbria, Italy. As we cannot work with a formal expression of the solution, in general, our fitting procedure was based on restricting first the range  of the various parameters by requiring good qualitative behaviors of the solutions (when compared with the real data), and then by performing  refined variations of the parameters trying each time to minimize the errors.

\begin{figure}
	\centering
	\includegraphics[width=\columnwidth]{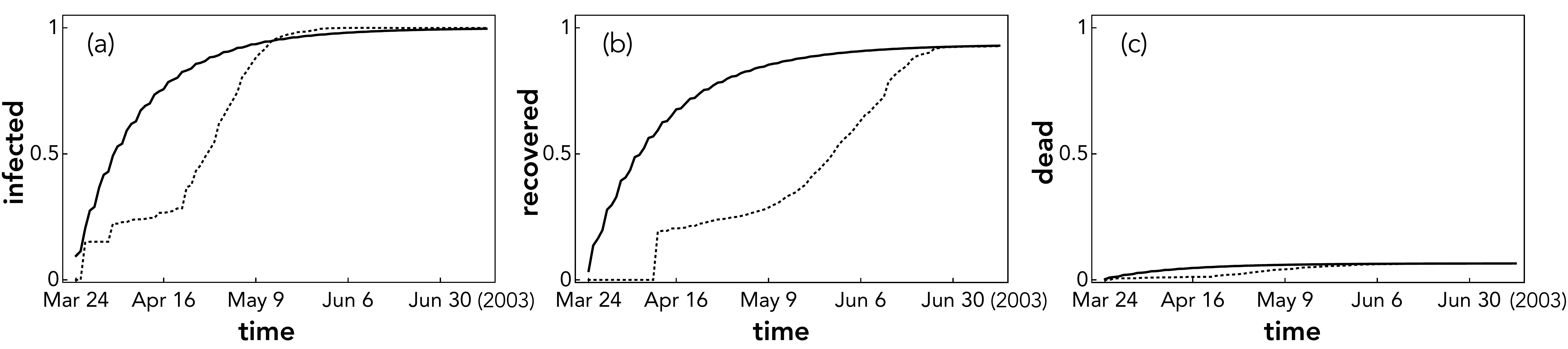}
	\caption{SARS 2003 in China. Real data (dotted) and model functions $P_j(t)$ (solid lines) {according to eqs.~\eqref{28}}. (a) cumulative infected, (b) cumulative recovered, (c) cumulative deaths. The initial state of $\Sc_{\cal P}$ is $\varphi_{0,0,0}$. Parameter values: $\Omega=1,\, N=1,\,  \omega_1=0.6,\,\omega_2=0.05,\,\omega_3=0.05,\, \sigma_{B,1}=0.15,\,\sigma_{B,2}=10^{-4},\, \sigma_{1,2}=1,\, \sigma_{1,3}=0.165$} 
	\label{fig:sars}
\end{figure}

\begin{figure}
	\centering
	\includegraphics[width=\columnwidth]{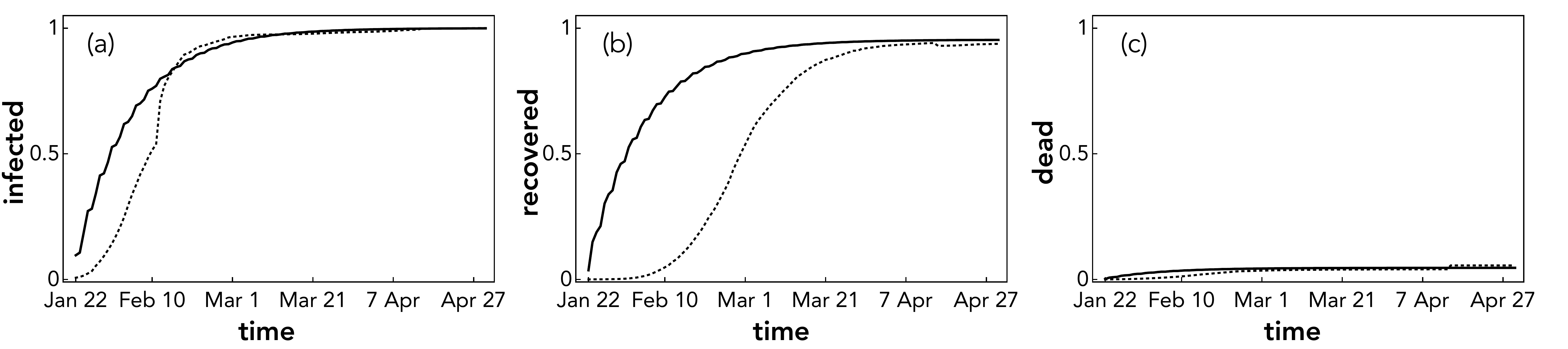}
	\caption{COVID-19 in China. Real data (dotted) and model functions $P_j(t)$ (solid lines) {according to eqs.~\eqref{28}}. (a) cumulative infected, (b) cumulative recovered, (c) cumulative deaths.  The initial state of $\Sc_{\cal P}$ is $\varphi_{0,0,0}$.Parameter values: $\Omega=4.5,\, N=1,\,  \omega_1=\omega_2=\omega_3=0,\, \sigma_{B,1}=0.32,\,\sigma_{B,2}=10^{-4},\, \sigma_{1,2}=1,\, \sigma_{1,3}=0.22$} 
	\label{fig:covid}
\end{figure}

\begin{figure}
	\centering
	\includegraphics[width=\columnwidth]{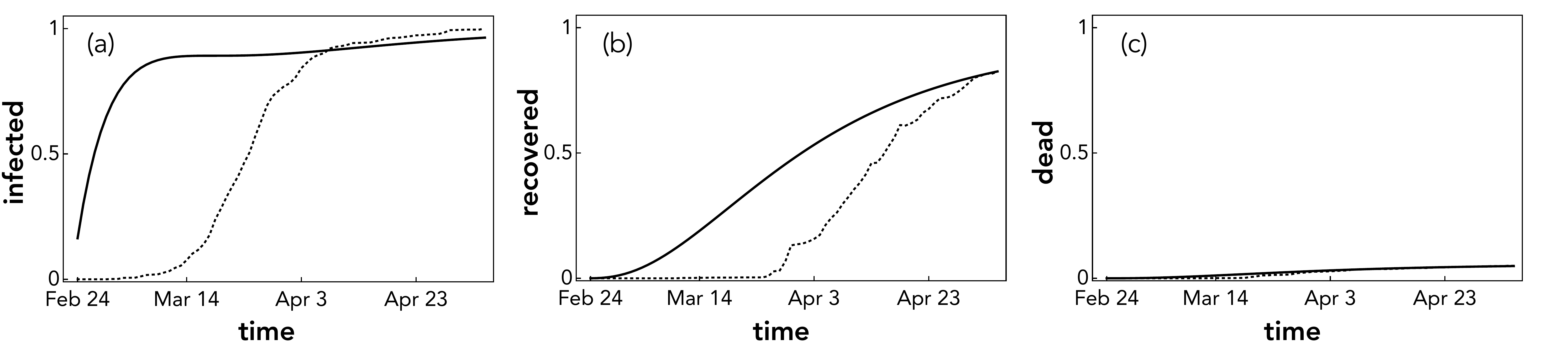}
	\caption{COVID-19 in Umbria. Real data (dotted) and model functions $P_j(t)$ (solid lines) {according to eqs.~\eqref{28}}. (a) cumulative infected, (b) cumulative recovered, (c) cumulative deaths.  The initial state of $\Sc_{\cal P}$ is $\varphi_{0,0,0}$. Parameter values: $\Omega=4.3,\, N=1,\,  \omega_1=\omega_2=\omega_3=0,\, \sigma_{B,1}=0.35,\,\sigma_{B,2}=0,\, \sigma_{1,2}=0.037,\, \sigma_{1,3}=0.009$} 
	\label{fig:covid_umbria}
\end{figure}

\begin{figure}
	\centering
	\includegraphics[width=\columnwidth]{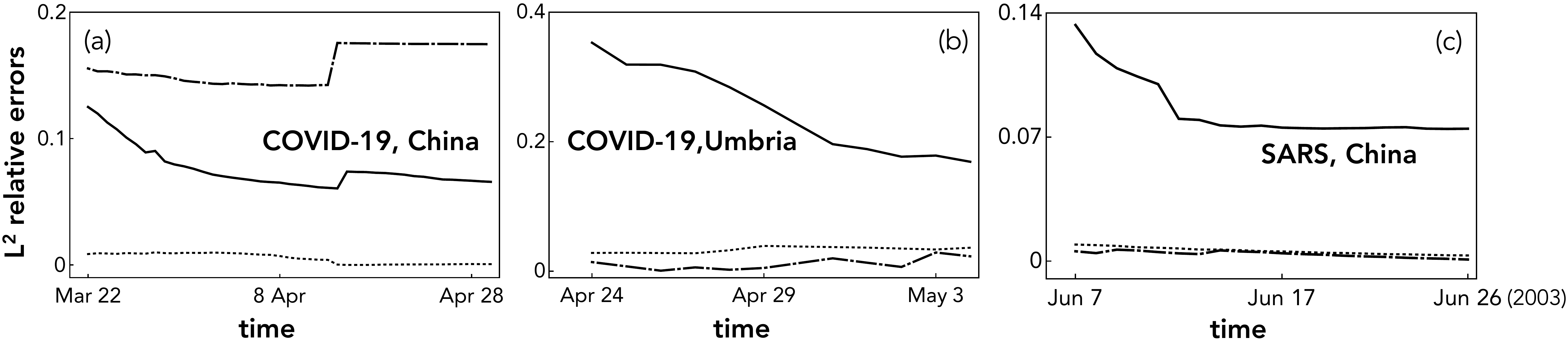}
	\caption{$L^2$ relative errors in the case of COVID-19 in China (a), in Umbria (b) and SARS in China (c). Cumulative infected (solid), cumulative recovered (dotted), cumulative deaths (dotted-dashed). Parameter values: same as Figures ~\ref{fig:sars},\ref{fig:covid},\ref{fig:covid_umbria}.} 
	\label{fig:err}
\end{figure}

All the real quantities are scaled with the final value of total confirmed cases, which in our cases are 5327 for the SARS, 80928 for the Coronavirus in China, and 1400 for the Coronavirus in Umbria, Italy\footnote{Final values used for the coronavirus are those at the moment of writing this paper.}.
Data are shown in Figure~\ref{fig:sars} {relative to} the period March-July 2003 for SARS epidemic, that is from the first reported infection up to plateau. Similar plots are shown in Figure~\ref{fig:covid} and~\ref{fig:covid_umbria} for Coronavirus in China and Umbria, respectively. All the initial densities are equal to zero. After the initial transient in which infection rapidly growths, and in which the model reproduces the real data only at a qualitatively level, we end with no increments and our model captures very well the final stage of the outbreak.  { In Figure~\ref{fig:err} we show some performance-metric results showing the $L^2$ relative errors (root mean squared errors) after the transient in the last time range of the analyzed periods. As one can see, the errors are in general very small, especially for the fitted recovered and dead, showing the good asymptotic predictability of our model}.


\subsection{Lockdown measures}\label{sectlockmea}
We consider now a simple variation of the model in which we adaptively change the parameters of the Hamiltonian to correct the Heisenberg dynamics to model the introduction of some  countermeasures  meant to fight the spreading of the disease. In particular, following the basic ideas of the $(H,\rho)$-induced dynamics introduced in \cite{fffr} and used in several other applications, \cite{bagbook2}, we change the value of the parameters 
(in our case just one)
of the Hamiltonian at some specific time, mimicking in this way the occurence of some check in the real-life situation. This induces a sort of discrete time dependence for the Hamiltonian.

To be more specific, we suppose that at the time  $\Delta T$ the parameter $N$ in \eqref{25} is adjusted to mimic the effects of some social rules, imposed by the the 
government,
which limit the diffusion of the disease. We recall that, as suggested by (\ref{212}), $N$ is a measure of the density of the exposed people, i.e. the reservoir, so that decreasing (resp. increasing)  $N$ limits (resp. eases) the infection process: in particular, decreasing $N$, we are modelling what has been called {\em social distance}.

Starting with an initial state $\Psi_0=\varphi_{0,0,0}$, we let the system evolve up to $\Delta T$, and we compute the densities using \eqref{28}.
Then we change the value of $N$, taking a lower value to mimic the beginning of a lockdown, and we start the new dynamics with a new initial state $$\Psi_{\textrm{new}}=\sum_{n_1,n_2,n_3}\alpha_{n_1,n_2,n_3}\varphi_{n_1,n_2,n_3},$$ which is constructed by requiring the continuity of the  densities (not their differentiability in general).
The  form of this state is obtained by choosing  the coefficients of $\Psi_{\textrm{new}}$ as follow:
\bea
\alpha_{1,0,0}&=&\sqrt{P_{1}(\Delta T)-P_{2}(\Delta T)-P_{3}(\Delta T)},\label{31}\\
\alpha_{1,1,0}&=&\sqrt{P_{2}(\Delta T)},\label{32}\\
\alpha_{1,0,1}&=&\sqrt{P_{3}(\Delta T)},\label{33}\\
\alpha_{0,0,0}&=&\sqrt{1-P_{1}(\Delta T)},\label{34}
\ena
while the other coefficients are set to 0.
Notice that  \eqref{31}-\eqref{34} ensures the continuity of the densities at $\Delta T$, and \eqref{34}, in particular, ensures the normalization of $\Psi_{\textrm{new}}$ to 1. The requirement that the other coefficients are 0 simply follows from our interpretation of the vectors of the basis $\F_\varphi$: it is not possible to have dead, or recovered, in absence of infected. This is the reason why we impose that $\alpha_{0,n_2,n_3}=0$, if $n_2$ or $n_3$, or both, are different from zero.  Moreover, we also put $\alpha_{1,1,1}=0$ since there is no possibility to have the same number of infected, recovered and dead. Of course there are other forms of the coefficients which ensure the continuity of the densities, but our choice works already rather well, and it is easy to implement. It might be useful to notice that the problem of fixing the coefficients in $\Psi_{\textrm{new}}$ would be authomatically solved if we were able to work in the Schr\"odinger, rather than in the Heisenberg, representation. But this is not easy at all, for our Hamiltonian.

\begin{figure}
	\centering
	\includegraphics[width=\columnwidth]{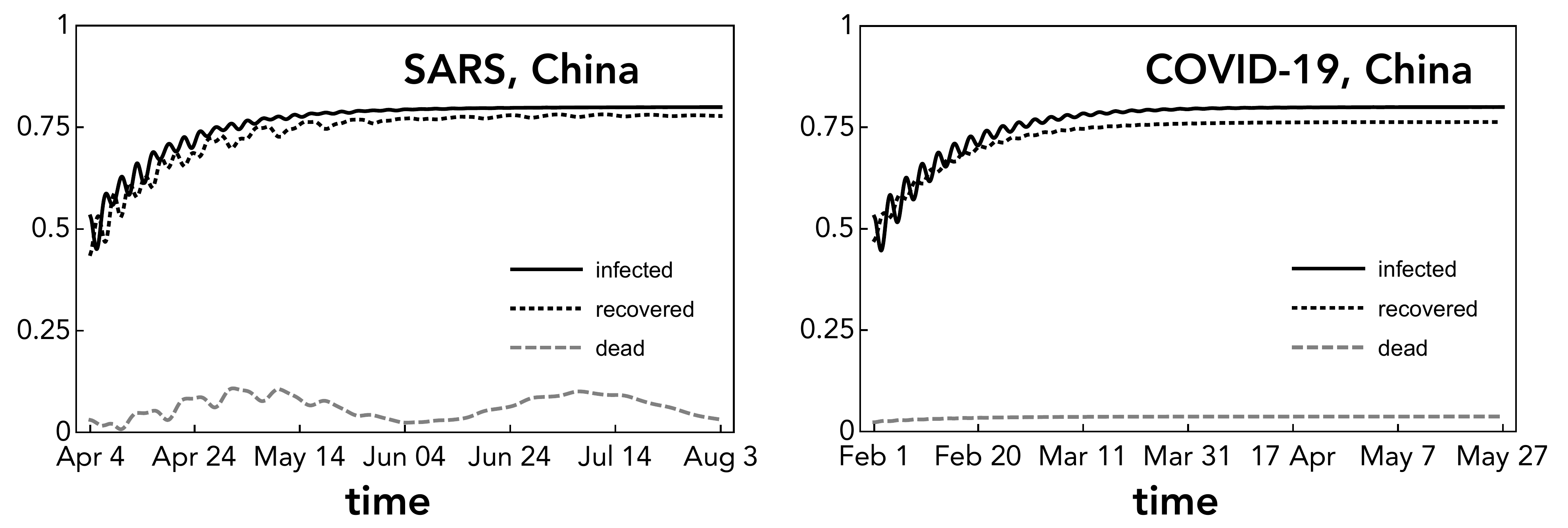}
	\caption{Behavior of the model functions $P_j(t)$ {in eqs.~\eqref{28}} for SARS (left) and COVID-19 (right), assuming a lockdown at the 10th day from the beginning of the spreading
		 Parameter values: same as Figures~\ref{fig:sars} and \ref{fig:covid} with $\Delta T=10$ and $N=0.8$. One can see that applying such a measure reduces in principle the asymptotic values of all populations.} 
	\label{fig:rules}
\end{figure}

\begin{figure}
	\centering
	\includegraphics[width=\columnwidth]{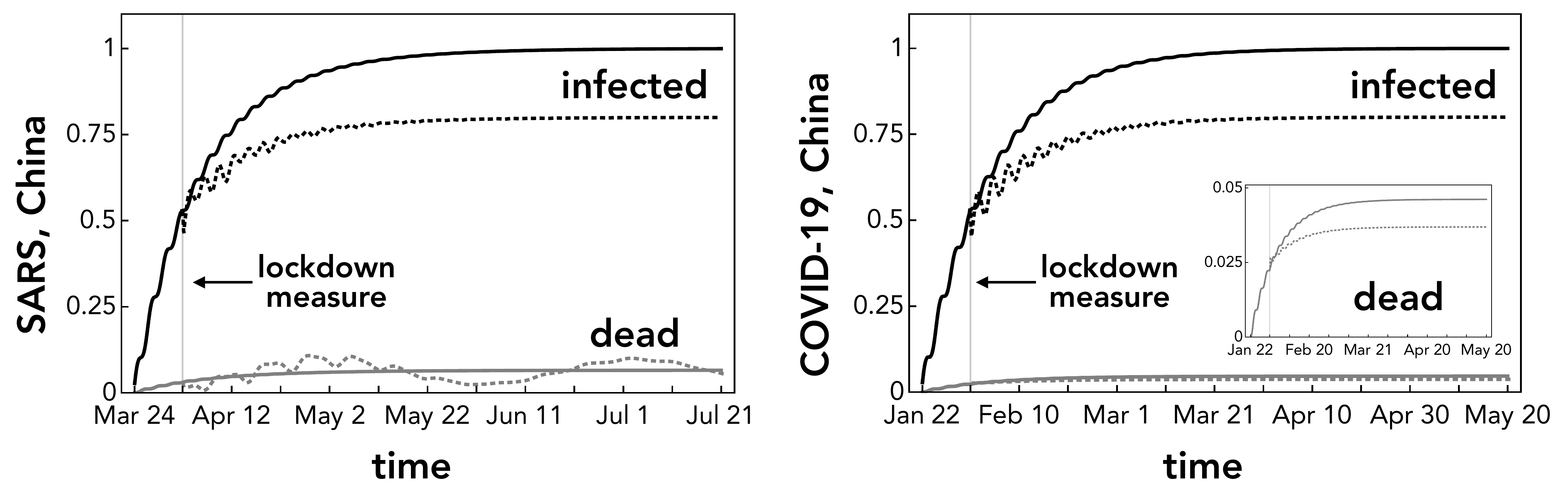}
	\caption{Comparison of the densities of infected (black curves) and dead (gray curves) for SARS (left) and COVID-19 (right), assuming a lockdown measure at the 10th day from the beginning of the spreading. Solid curves densities without lockdown measures, dotted curves densities with measures. In the inset of the
		right figure the magnification of the evolution of the dead densities.
		Parameter values: same as Figures~\ref{fig:sars} and \ref{fig:covid}, with $\Delta T=10$ and $N=0.8$.  } 
	\label{fig:dualplot}
\end{figure}

In Figure \ref{fig:rules}, we present  the results concerning  SARS and COVID-19, using the same parameters as reported in Figure \ref{fig:sars} and Figure \ref{fig:covid}, respectively, and with $\Delta T=10$ and $N=0.8$: this is like saying that, after 10 days, the lockdown measures lowered the {\em normal } density of the people from 1 to 0.8. A direct comparison of the density functions with and without the applied measures is shown in Figure \ref{fig:dualplot}.
Of course different values of $\Delta T$ and lower values of $N$ could be chosen, producing quantitative difference in the results, but with similar qualitative outcomes (i.e. reduction of the infection).
We can observe here that oscillations in the transient  are evident, and disappear with the exception of the dead density in the SARS case: in this case  the asymptotic behavior of the density is an oscillating function (not fully shown in the figure) with period $T\approx90$ and mean value $\bar P_3=0.036$. It is interesting to stress that this value is much lower than the asymptotic value $P_3(\infty)=0.065$ obtained without any lockdown measure.  Therefore, the procedure of imposing a lockdown at a certain time via a certain social distance, is able, in the model, to reduce significantly the asymptotic number of infected, recovered and dead individuals, in agreement with what observed in real life.

\section{Conclusions}\label{sec:conclo}

In this paper we propose a  model for the spreading of infection diseases which is different from those based on differential equations or stochastic approaches. We adopt an operatorial approach inspired by quantum mechanics and based on creation and annihilation operators. This is because ladder operators allow to encode, in a simple and elegant way, real interactions among different individuals through Hamiltonian interaction terms and to consider one population as a sort of \textit{reservoir} for the others. The application to the spreading of SARS in 2003 in China and of COVID-19 in China and in the Italian region of Umbria shows how such a model is able to  capture well the asymptotic behavior of the spreading. In addition, imposing lockdown measures by reducing a model parameter at a precise instant of time, we have shown that it is possible to reduce significantly the number of infected individuals, and of the recovered and dead as a consequence. Of course the model proposed here is only a first, but promising, proposal, where a lot of improvements could be added to make it more precise and, possibly, useful. Also, it would be interesting to try to improve the short time behavior, to better reproduce what happens in real life. These are only part of our future projects.

%
%
%

\section*{Appendix 1:  Few results on the CAR}
We say that a set of operators
$\{a_\ell,\,a_\ell^\dagger, \ell=1,2,\ldots,L\}$ acting on an Hilbert space $\Hil$ satisfy the CAR if the conditions $$ \{a_\ell,a_n^\dagger\}=\delta_{\ell n}\1,\hspace{8mm}
\{a_\ell,a_n\}=\{a_\ell^\dagger,a_n^\dagger\}=0 $$ hold true for all $\ell,n=1,2,\ldots,L$. Here, $\1$ is the identity operator
and $\{x,y\}:=xy+yx$ is the {\em anticommutator} of $x$ and $y$. These operators are used to describe $L$ different \emph{modes} of fermions. From these
operators we can construct $\hat n_\ell=a_\ell^\dagger a_\ell$ and $\hat N=\sum_{\ell=1}^L \hat n_\ell$, which are both self--adjoint. In
particular, $\hat n_\ell$ is the \emph{number operator} for the $\ell$--th mode, while $\hat N$ is the \emph{global number operator}.

An orthonormal basis for $\Hil$  is constructed as follows: we introduce the \emph{vacuum} of the theory, that is a vector $\varphi_{\bf 0}$
which is annihilated by all the operators $a_\ell$: $a_\ell\varphi_{\bf 0}=0$ for all $\ell=1,2,\ldots,L$. Such a non zero vector surely exists. Then we act on $\varphi_{\bf 0}$
with the  operators $a_\ell^\dagger$ (but not with higher powers, since these powers are simply zero!): $$
\varphi_{n_1,n_2,\ldots,n_L}:=(a_1^\dagger)^{n_1}(a_2^\dagger)^{n_2}\cdots (a_L^\dagger)^{n_L}\varphi_{\bf 0}, $$ $n_\ell=0,1$ for
all $\ell$. These vectors form the orthonormal basis we were looking for, and are eigenstates of both $\hat n_\ell$ and $\hat N$: $\hat
n_\ell\varphi_{n_1,n_2,\ldots,n_L}=n_\ell\varphi_{n_1,n_2,\ldots,n_L}$ and $\hat N\varphi_{n_1,n_2,\ldots,n_L}=N\varphi_{n_1,n_2,\ldots,n_L},$
where $N=\sum_{\ell=1}^Ln_\ell$. Moreover, using the  CAR, we deduce that $$\hat
n_\ell\left(a_\ell\varphi_{n_1,n_2,\ldots,n_L}\right)=(n_\ell-1)(a_\ell\varphi_{n_1,n_2,\ldots,n_L})$$ and $$\hat
n_\ell\left(a_\ell^\dagger\varphi_{n_1,n_2,\ldots,n_L}\right)=(n_\ell+1)(a_l^\dagger\varphi_{n_1,n_2,\ldots,n_L}),$$ for all $\ell$. Then
$a_\ell$ and $a_\ell^\dagger$ are  called the
\emph{annihilation} and the \emph{creation} operators. Notice that, in some sense, $a_\ell^\dagger$ is {\bf also} an annihilation operator since,
acting on a state with $n_\ell=1$, we destroy that state.

Of course, $\Hil$ has a finite dimension. In particular,
for just one mode of fermions, $dim(\Hil)=2$.

\section*{Appendix 2: few closed formulas}

There are cases in which the functions in (\ref{28}) can be computed analytically and assume a reasonable closed form. For instance, if we take $\omega_{j}=0$ for $j=1,2,3$ and $\sigma_{B,2}=0$, then, by setting 

\begin{equation}
\nu = \sqrt{\pi ^2 \sigma _{B,1}^4-4 \Omega ^2 \left(\sigma _{1,2}^2+\sigma _{1,3}^2\right)}
\end{equation}
we have that

\begin{eqnarray}
P_1(t)
& = & -\frac{N e^{-\frac{t \left(2 \pi  \sigma _{B,1}^2+\nu \right)}{\Omega }}\left(\left(8 \Omega ^2 \left(\sigma _{1,2}^2+\sigma _{1,3}^2\right)-2 \pi ^2 \sigma _{B,1}^4\right) e^{\frac{t \left(2 \pi  \sigma _{B,1}^2+\nu \right)}{\Omega }}+\pi 
	\sigma _{B,1}^2 \left(\pi  \sigma _{B,1}^2+\nu \right) e^{\frac{\pi  t \sigma _{B,1}^2}{\Omega }}\right)}{2 \nu^2} \nonumber \\
& &  -\frac{N e^{-\frac{t \left(2 \pi  \sigma _{B,1}^2+\nu \right)}{\Omega }} \left( 
	\pi  \sigma _{B,1}^2 \left(\pi  \sigma _{B,1}^2-\nu \right) e^{\frac{t \left(\pi  \sigma _{B,1}^2+2 \nu \right)}{\Omega }}-8 \Omega ^2 \left(\sigma _{1,2}^2+\sigma
	_{1,3}^2\right) e^{\frac{t \left(\pi  \sigma _{B,1}^2+\nu \right)}{\Omega }} \right) }{2 \nu^2} 
\end{eqnarray}

\begin{eqnarray}
P_2(t)
& = & -\frac{N \sigma _{1,2}^2 e^{-\frac{t \left(\pi  \sigma _{B,1}^2+\nu \right)}{\Omega }} \left(8 \Omega ^2 \sigma _{1,2}^2 e^{\frac{\nu  t}{\Omega }} \left(e^{\frac{\pi  t \sigma _{B,1}^2}{\Omega }}-1\right)+8 \Omega ^2 \sigma _{1,3}^2 e^{\frac{\nu 
			t}{\Omega }} \left(e^{\frac{\pi  t \sigma _{B,1}^2}{\Omega }}-1\right)\right)}{2 \left(\sigma _{1,2}^2+\sigma _{1,3}^2\right) \nu^2}  \nonumber \\
& &   -\frac{N \sigma _{1,2}^2 e^{-\frac{t \left(\pi  \sigma _{B,1}^2+\nu \right)}{\Omega }} \pi  \sigma _{B,1}^2 \left(\pi  \sigma _{B,1}^2 \left(-2 e^{\frac{t \left(\pi  \sigma _{B,1}^2+\nu \right)}{\Omega }}+e^{\frac{2 \nu  t}{\Omega }}+1\right)+\nu 
	\left(e^{\frac{2 \nu  t}{\Omega }}-1\right)\right)}{2 \left(\sigma _{1,2}^2+\sigma _{1,3}^2\right)\nu^2}\nonumber\\
\end{eqnarray}

and

\begin{equation}
P_3(t) = \frac{\sigma _{1,3}^2}{\sigma _{1,2}^2}P_2(t)
\end{equation}

from which  the results in \eqref{limits} follow.

\end{document}